\documentclass[12pt]{iopart}
\usepackage{epsf,epsfig,amssymb,graphicx,times}
\begin{document}
\title[Gaussian states entanglement and applicability over fading channels]{Entanglement of Gaussian states and the applicability to quantum key distribution over fading channels}
\author{Vladyslav C Usenko$^{1,2}$, Bettina Heim$^{3,4,5}$, Christian Peuntinger$^{3,4}$, Christoffer Wittmann$^{3,4}$, Christoph Marquardt$^{3,4}$, Gerd Leuchs$^{3,4}$ and Radim Filip$^1$}
\address{$^1$ Department of Optics, Palack\' y University, 17. listopadu 12,  771~46 Olomouc, Czech Republic}
\address{$^2$ Bogolyubov Institute for Theoretical Physics of National Academy of Sciences,
Metrolohichna st. 14-b, 03680, Kiev, Ukraine}
\address{$^3$ Max Planck Institute for the Science of Light, Guenther-Scharowsky-Str. 1/Bldg 24, 91058 Erlangen, Germany}
\address{$^4$ Institute of Optics, Information and Photonics, Friedrich-Alexander University of Erlangen-Nuremberg (FAU), Staudtstr. 7/B2, 91058 Erlangen, Germany}
\address{$^5$ Erlangen Graduate School in Advanced Optical Technologies (SAOT), FAU, Paul-Gordan-Str. 6, 91052 Erlangen, Germany}
\eads{\mailto{usenko@optics.upol.cz}, \mailto{bettina.heim@mpl.mpg.de}}
\date{\today}
\begin{abstract}
Entanglement properties of Gaussian states of light as well as the security of continuous variable quantum key distribution with Gaussian states in free-space fading channels are studied. These qualities are shown to be sensitive to the statistical properties of the transmittance distribution in the cases when entanglement is strong or when channel excess noise is present. Fading, i.e. transmission fluctuations, caused by beam wandering due to atmospheric turbulence, is a frequent challenge in free-space communication. We introduce a method of fading discrimination and subsequent post-selection of the corresponding sub-states and show that it can improve 
the entanglement resource and restore the security of the key distribution over a realistic fading link. Furthermore, the optimal post-selection
strategy in combination with an optimized entangled resource is shown to drastically increase the protocol robustness to excess noise, which is confirmed for experimentally measured fading channel characteristics. The stability of the result against finite data ensemble size and imperfect channel
estimation is also addressed.
\end{abstract}
\maketitle

\section{Introduction}
Entanglement in quantum states of light is well known to be an essential resource for quantum information processing, 
in particular for quantum key distribution (QKD) \cite{qkd1,qkd2}, which is up to date the most developed application in the field of quantum information. As an alternative to the first proposals \cite{dv_theor1,dv_theor2} and numerous QKD implementations based on discrete variable coding (e.g. the recent ones in fiber \cite{dv_exp1,dv_exp2} and open-space \cite{dv_exp3} channels), protocols for QKD were developed using a continuous variable (CV) approach \cite{cv}, aimed at improving the robustness and applicability of quantum information protocols.
The implementation of CV QKD protocols in their optical realization is well studied in the case of optical fiber channels. 
In particular, the security of CV QKD with Gaussian states was proven for any channel with 
pure constant attenuation and, to some extent, for noisy channels with constant attenuation \cite{cvqkd1,cvqkd2,cvqkd3,cvqkd31,cvqkd4,cvqkd41}. 
However, Gaussian protocols need additional analysis when atmospheric
channels are considered, since such channels show a prominent, temporal evolution of the channel transmission. Indeed, in free-space links fixed attenuation is 
never the case, as the transmittance fluctuates due to atmospheric 
effects. Such fading channels are characterized by a distribution of transmittance values. At the same time, 
atmospheric links are of utmost importance, providing the possibility for short- and mid-range communication without strong requirements to the infrastructure and, moreover, the possibility for truly long-distance communication, e.g. over satellite connections. 

In the case of discrete variable states and protocols, the impact of atmospheric channels on qubit-based QKD was actively investigated, for example, in \cite{dvatm,dvatm1,dvatm2,dvatm3,dvatm4}. The most negative effect of the free-space channel in this case is the background light (in the CV-based implementations it is effectively filtered out by the homodyne detection \cite{B4}). The impact of fading channels on the discrete variable protocols is similar to that of a constant attenuation, and results in the decrease of the communication rate (which can also be partly compensated by filtering \cite{dvatm5}), but can preserve the entanglement properties upon postselected measurements \cite{polentfad}. However, the effect of fading on CV quantum states is essentially different from the typical fixed loss and additive Gaussian noise present in fiber-type channels. The fading noise is multiplicative, which on the one hand quickly destroys the Gaussian quantum features such as entanglement and purity but on the other hand offers different possibilities to recover them \cite{distillation,filtering,distillation1,distillation2}. The noise due to atmospheric fluctuations is relatively slow (typically several kHz \cite{zeilinger}) compared to the achievable homodyne detection rates (hundreds of MHz \cite{bellini}), 
which means that it can be partially estimated using a strong optical beam, when entanglement properties are analyzed. In the case of QKD, the eavesdropping scenario must be considered. Therefore, the fading properties must be identified using the same alphabet of quantum states as used for communication. 

In the present paper, we analyze the effect of an atmospheric fading channel on the entanglement and security properties of protocols using Gaussian quantum states of 
light and consider the use of post-selection (PS) of sub-channels with the aim of improving the entangled resource and restore 
the security of the CV QKD protocol in case it was broken by the atmospheric fluctuations. We demonstrate the negative impact of fading on Gaussian entanglement and security, considering a general class of fading channels as well as a particular theoretical model of fading caused by beam wandering \cite{rostock} and show the possibility to restore entanglement and security by PS of sub-channels. Thus, based on the knowledge of the overall channel properties obtained from estimation, a higher key rate or stronger entanglement can be shared by the proper selection of sub-channels with less fading. Such a method presents a feasible alternative to the experimentally challenging state distillation \cite{distillation}. Moreover, we analyze the PS method considering experimentally measured transmission statistics of a real free-space channel and find an increase of the secret key rate in such a realistic scenario.

The paper is organized as follows. In section \ref{fading}, we describe fading channels in general and show the negative effect of fading on Gaussian states entanglement and the possible appliation to CV QKD. In section \ref{beam-wandering}, we theoretically consider the particular case of fading channels, where transmittance fluctuations are caused by beam wandering, showing the effect of fading on entanglement and security. We introduce PS, which we show capable of restoring entanglement of the states and security of the respective Gaussian protocol. In section \ref{real-channel}, we apply our analysis to a real fading channel, transmittance distribution of which the was measured experimentally. In section \ref{finite-size}, we numerically address the finite-size effects of the limited data ensembles and imperfect estimation on the PS of sub-channels, showing the stability of our results in the finite-size context. 

\section{Fading channels}
\label{fading}
\begin{figure}[h]
\centerline{\includegraphics[width=0.7\textwidth]{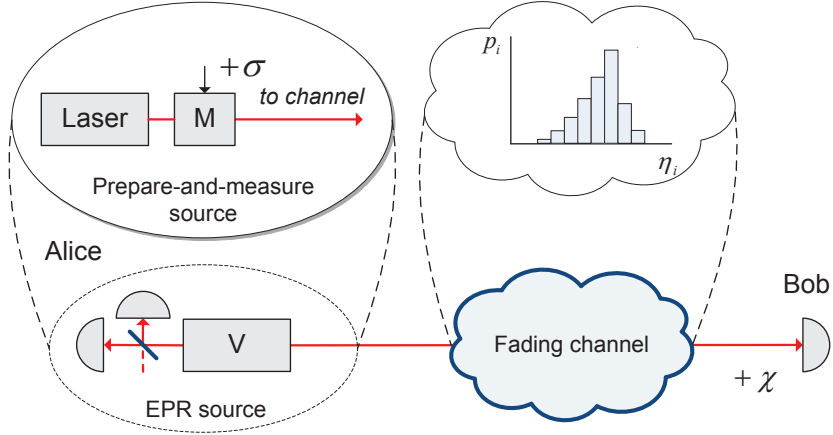}}
\caption{EPR entanglement-based QKD over a fading channel. Alice and Bob share an entangled state 
of variance $V$; one of the modes is transmitted to Bob through a fading channel described by a distribution of transmittance 
values $\{\eta_i\}$. While Alice directly applies heterodyne measurements, Bob makes homodyne measurements allowing him to 
characterize the channel and estimate excess noise. Left inset: the equivalent prepare-and-measure scheme. Alice generates a coherent state, displaces it in phase-space with a variance of $\sigma=V-1$ using the modulator M, and sends it to Bob. Right inset: illustrative discretized transmittance distribution
of a fading channel. 
\label{fading_channel}}
\end{figure}

A quantum communication scheme over a fading channel is presented in figure \ref{fading_channel}. Alice and Bob share an entangled two-mode squeezed vacuum \cite{cv} state which is characterized by the quadrature variance $V$ of each of its modes. Half of the entangled state is sent through a fading channel to Bob. Bob makes a homodyne measurement of the amplitude or
phase quadrature, and Alice measures both quadratures in a balanced heterodyne scenario. Such an entanglement-based CV QKD scheme is equivalent to a 
prepare-and-measure scheme (see figure \ref{fading_channel}, left inset), in which Alice generates a coherent state, applies random Gaussian displacement
of variance $\sigma=V-1$ to the generated state and sends it through a fading channel to Bob, who then also makes a homodyne measurement. Note, that while the entangled state 
variance $V$ is related to the nonclassicality of a pure state (being $V=\cosh{2r}$, where $r$ is the squeezing parameter) and high values (tens of shot-noise units (SNU)) are experimentally challenging to achieve, the displacement variance $\sigma$ in the prepare-and-measure scenario is the result of a feasible modulation and can reach high values (up to hundreds of SNU). Thus, despite the simple equivalence between those two implementations, we stick to the limited values of $V$ when considering entanglement properties, but assume a wider region for $\sigma$ when estimating the security, as it can always be implemented with a high modulation variance in the prepare-and-measure scenario.

In our work, we deal with the Gaussian states of light. Such states can be explicitly described by the covariance matrices, i.e. matrices of the second moments of the quadrature operators, which we introduce through mode annihilation and creation operators as $x=a^\dag+a$ and $p=i(a^\dag-a)$. Thus, variance of the vacuum fluctuations is $1$.

Let the channel be described by a distribution of transmittance values $\{\eta_i\}$ with probabilities $\{p_i\}$. 
The evolution of a Gaussian state over a purely attenuating fading channel was shown to be governed by two 
parameters of the attenuation distribution - the mean value of transmittance $\langle \eta \rangle$ and the mean of the square root 
of transmittance $\langle \sqrt{\eta} \rangle$ \cite{distillation2}. Indeed, the covariance matrix of a two-mode 
squeezed vacuum state with variance $V\geqslant 1$
\begin{equation}
\label{inputstate}
\gamma_{AB} =
\left( \begin{array}{cc}
V\mathbb{I} & \sqrt{V^2-1}\sigma_z \\
\sqrt{V^2-1}\sigma_z & V\mathbb{I} 
\end{array} \right)
\end{equation} 
after a channel with transmittance $\eta_i$ in mode B is given by 

\begin{equation}
\label{ithchan}
\gamma^i_{AB} =
\left( \begin{array}{cc}
V\mathbb{I} & \sqrt{\eta_i} \sqrt{V^2-1}\sigma_z \\
\sqrt{\eta_i} \sqrt{V^2-1}\sigma_z & (V \eta_i + 1 - \eta_i + \chi)\mathbb{I}
\end{array} \right),
\end{equation} 

following the input-output relation for the quadrature vectors
$r=(x,p)^T$ of modes 1 (signal mode B in our case) and 2 (vacuum input mode of the purely lossy i-th sub-channel) in the form
\begin{eqnarray}
\label{inout}
\left( \begin{array}{c}
r_1 \\
r_2
\end{array} \right)_{out} =
\left( \begin{array}{cc}
\sqrt{\eta_i}\mathbb{I} & \sqrt{1-\eta_i}\mathbb{I} \\
-\sqrt{1-\eta_i}\mathbb{I} & \sqrt{\eta_i}\mathbb{I}
\end{array} \right)
\left( \begin{array}{c}
r_1 \\
r_2
\end{array} \right)_{in},
\end{eqnarray}
where $\sqrt{\eta_i}$ stands for the transmittance of a coupling beamsplitter, $\mathbb{I}=diag(1,1)$ is the unity matrix and $\sigma_z=diag(1,-1)$ is the Pauli matrix. We also assume the presence of excess noise $\chi$, which we later consider untrusted in the so-called "pessimistic scenario" (this can be detector noise or untrusted channel noise, scaled by the channel attenuation). 

The overall state after a fading channel is the mixture of states after individual sub-channels; thus the Wigner function of the state is the sum of Wigner functions of the states after sub-channels weithed by sub-channel probabilities \cite{distillation2}. The elements of the total covariance matrix after fading channel $\{\eta_i\}$ are then the convex sum of the moments given by ($\ref{ithchan}$) and the covariance matrix of the resulting mixed state after a fading channel $\{\eta_i\}$ in mode B is then given by
\begin{equation}
\label{evolvedstate}
\gamma'_{AB} =
\left( \begin{array}{cc}
V\mathbb{I} & \langle \sqrt{\eta} \rangle\sqrt{V^2-1}\sigma_z \\
\langle \sqrt{\eta} \rangle\sqrt{V^2-1}\sigma_z & (V\langle \eta \rangle + 1 - \langle \eta \rangle + \chi)\mathbb{I}
\end{array} \right).
\end{equation} 

Let us estimate the impact of a fading channel on the quantum properties of a state, propagating through such a channel and on the security 
of coherent state CV QKD. From the covariance matrix of the state 
after the fading channel (\ref{evolvedstate}) it is evident, that the noiseless fading channel can be considered as a non-fading 
channel with transmittance $\langle \sqrt{\eta} \rangle ^2$ and excess noise caused by fading
$\epsilon_{f}=Var(\sqrt{\eta})(V-1)$ so that variance of the mode B after the fading channel becomes
\begin{equation}
V_B'=\langle \sqrt{\eta} \rangle^2(V-1)+\epsilon_{f}+\chi+1,
\end{equation} 
where $Var(\sqrt{\eta})=\langle \eta \rangle - \langle \sqrt{\eta} \rangle ^2$. Thus, the effect of a fading channel in terms of the covariance matrix can be described as Gaussian state variance-dependent excess noise. Further we focus on the properties of the state, described by the covariance matrix, as they are crucial for Gaussian CV QKD protocols.

Evidently, fading reduces the purity of the initially pure Gaussian states compared to channels with fixed attenuation. The purity of a Gaussian two-mode state, described by the covariance matrix $\gamma_{AB}$, can be expressed as $p(\gamma_{AB})=1/\sqrt{Det \gamma_{AB}}$ \cite{measures} (thus having the meaning of the Gaussian mixedness of the state), $Det \gamma_{AB}$ being the determinant of the covariance matrix. In case of the fading channel, acting on the state (\ref{inputstate}), the purity reads
\begin{equation}
p(\gamma'_{AB})=\frac{1}{Var(\sqrt{\eta})V(V-1)+V(1-\langle\sqrt{\eta}\rangle^2)+\langle\sqrt{\eta}\rangle^2}.
\end{equation}
As fading increases (which means increase of the variance $Var(\sqrt{\eta})$), the purity is reduced, while for an arbitrarily strong fading $Var(\sqrt{\eta}) = 1/4$ (corresponding to equiprobable transmittance values of 0 and 1) the purity is given by $p(\gamma_{AB})=4/(V+1)^2$, which is degraded with increasing $V$ compared to any channel with fixed transmission. In the opposite limit of two uncorrelated vacuum modes $V=1$ the state is always pure with $p=1$.

\begin{figure}[h]
\centerline{\begin{tabular}{ll}
\includegraphics[width=0.35\textwidth]{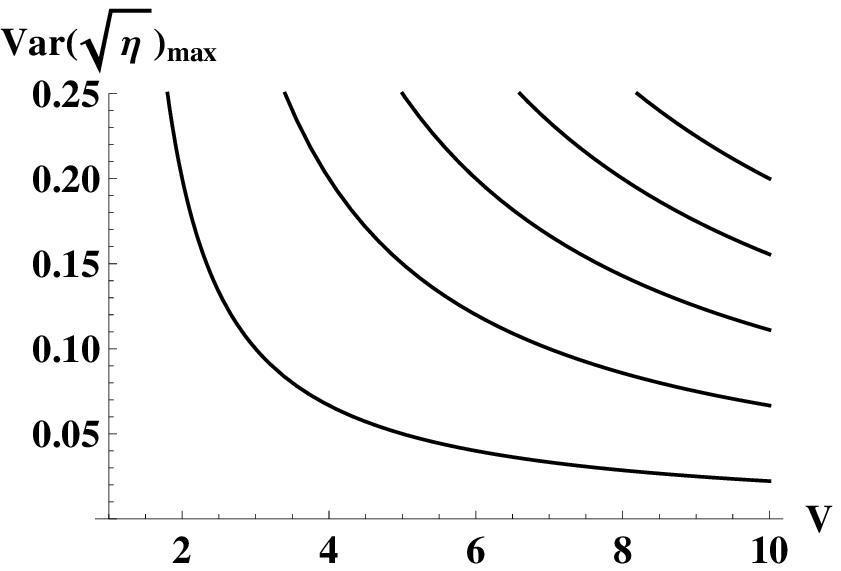}
\includegraphics[width=0.35\textwidth]{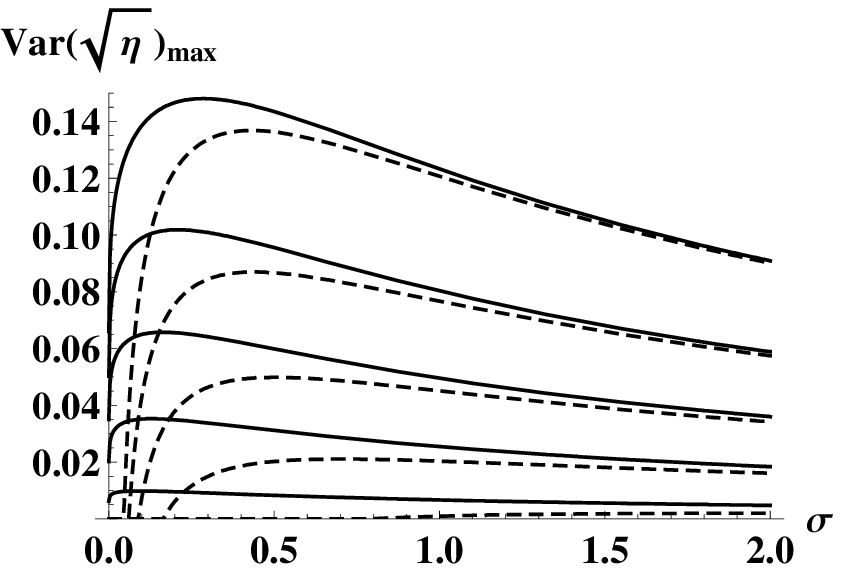}	
\end{tabular}}
\caption{Left: maximal entanglement-preserving $Var(\sqrt{\eta})$ (in terms of logarithmic negativity) versus the state variance $V$ for different values of $\langle \sqrt{\eta} \rangle^2$ (from top to bottom: $0.9,0.7,0.5,0.3,0.1$) in the absence of excess noise (which otherwise does not significantly affect the obtained dependencies). Right: maximal security-preserving $Var(\sqrt{\eta})$ (in the case of collective attacks) versus Gaussian modulation variance $\sigma=V-1$ for different values of $\langle \sqrt{\eta} \rangle^2$ (from top to bottom: $0.9,0.7,0.5,0.3,0.1$) in the absence of untrusted noise (solid lines) and upon 
untrusted detection noise $\chi=0.012$ (dashed lines). 
\label{maxvar}}
\end{figure}

Similarly, fading reduces Gaussian entanglement, which we consider in terms of log-negativity \cite{vidal} $E_{LN}(\gamma)=max[0,-ln(\tilde\lambda_{-})]$, where $\tilde\lambda_{-}$ is the smallest symplectic eigenvalue of the partially transposed state of the Gaussian state described by the covariance matrix $\gamma$ \cite{measures}. Gaussian entanglement of the state described by the covariance matrix (\ref{evolvedstate}) is then broken given a variance $V$ by the value 
of $Var(\sqrt{\eta})_{max,ent}=2\langle\sqrt{\eta}\rangle^2/(V-1)$. If detector or other excess noise is present, then the maximum entanglement-breaking fading variance is
\begin{equation}
\label{maxvarent}
Var(\sqrt{\eta})_{max,ent}=\frac{
2(\langle\sqrt{\eta}\rangle^2-1)-\chi+\sqrt{4(1+\langle\sqrt{\eta}\rangle^2)^2+\chi^2}
}{
2(V-1)
},
\end{equation}
where $\chi$ is the excess noise after channel, assumed to be independent from fading.
For high state variance $V \gg 1$ the maximum tolerable $Var(\sqrt{\eta})_{max,ent}$ tends to 0, which means that even small fading is Gaussian entanglement breaking. At the same time, when the variance is low ($V \to 1$), entanglement is preserved for any fading. The typical behavior of the fading variance threshold with respect to state variance is given in figure \ref{maxvar} (left) for different mean transmittance values. Evidently, the robustness of entanglement to fading is drastically decreased with an increase of state variance, while additional detection (or channel) excess noise of reasonably small amounts does not play a significant role.

Let us now consider the impact of a fading channel on the security of a protocol based on Gaussian states. We refer to a certain CV QKD coherent state protocol \cite{cvqkd1}, whose security analysis in the general case is based on virtual Gaussian entanglement \cite{equiv}. Robustness of the protocol against collective attacks (which also implies security against the most general coherent attacks \cite{cohatt}) was shown using the extremality of Gaussian states \cite{extremality} and the subsequent optimality of Gaussian attacks \cite{security1,security2}. This enables a security analysis based solely on the covariance matrix description of the states. However, this implies 
the necessity for the trusted parties to know the channel characteristics in order to explicitly derive the upper bound on the knowledge about the key bits 
accessible to a potential eavesdropper. This and the further assumptions required to analyze the security of the coherent state protocol over a fading channel are explicitly given below.

Firstly, we make the standard CV QKD assumption that the trusted parties are able to estimate the channel transmittance $\eta$ and check its stability during the transmission of data, which will contribute to the secure key (we refer to a given relatively stable transmittance as a sub-channel). This is experimentally feasible, as the typical rate of atmospheric channel fluctuations is in the order of kHz, while the modulation and detection rate is typically in the order of several MHz, i.e. at least thousands of signal or probe states can be transmitted during the stability time of the fading atmospheric channel. Thus, each measurement result in our case can be reliably attributed to the respective sub-channel within the given transmittance window, 
over which the given signal state was transmitted. 


Secondly, in order to make our analysis clear and focus on the discussed effects, we assume that the entangled (or prepared) states are pure. Consideration of the realistic noisy sources will be the subject of further research. For the same reason and with the same perspectives we assume perfect data post-processing. 

Thirdly, we perform the analysis in the asymptotic limit, but finish the paper with the numerical estimation of the finite data ensemble size effect, as it could be crucial in the conditions of the ensemble reduction due to PS. Similarly we address the effects of imperfect channel loss estimation on our results.

Relying on the above-given assumptions we consider in detail the effect of fading channels and the possibility to improve entanglement of the Gaussian states and security of the respective CV QKD protocols by PS.

A lower bound on the key rate in the case of the most effective collective eavesdropping attacks in the reverse reconciliation scenario 
is well known to be $K=I_{AB}-\chi_{BE}$ \cite{security1,security2}, where $\chi_{BE}$ is the Holevo quantity, which gives an upper bound for the 
leaked information, and is calculated trough the von Neumann entropy of the eavesdropper state 
before the measurement on mode $B$ and the state conditioned by the result of the measurement. 

The mutual information $I_{AB}$ between the trusted parties can be expressed through variances and conditional variances of modes $A$ and $B$ as $I_{AB}=(1/2)\log{(V_A^M/V_{A|B}^M})$ after heterodyne measurement at Alice (corresponding to conditional coherent state preparation). In this case Alice is measuring the observable $x_A^M=(x_A+x_0)/\sqrt{2}$ ($x_M$ being the quadrature of mode A prior to heterodyne and $x_0$ being the quadrature of the vacuum input of the free port of a heterodyne detector), and mutual information reads  
\begin{equation}
I_{AB}=\frac{1}{2}\log{
\frac{1}
{
1-\frac{\langle\sqrt{\eta}\rangle^2\sigma}{1+\langle\eta\rangle\sigma+\chi}
}
}.
\end{equation}
The Holevo quantity, under the assumption that Eve is able to purify the state \cite{cvqkd4}, is calculated through the symplectic eigenvalues $\lambda_{1,2}$ of the covariance matrix (\ref{evolvedstate}) after the channel and the symplectic eigenvalue $\lambda_3$ of the conditional matrix 
\begin{equation}
\gamma_A^{x_B}=\gamma_A-\sigma_{AB}(X \gamma_B X)^{MP}\sigma_{AB}^T,
\end{equation}
where $\gamma_A$, $\gamma_B$ are the matrices, describing the modes A and B individually and $\sigma_{AB}$ is the matrix, which characterizes correlations between the modes A and B. These matrices all together construct the covariance matrix (\ref{evolvedstate}). MP stands for Moore Penrose inverse of a matrix (also known as pseudoinverse), and
\begin{equation}
X =
\left( \begin{array}{cc}
1 & 0 \\
0 & 0
\end{array} \right).
\end{equation}
Here with no loss of generality we assume that the x-quadrature is measured by Bob.

The Holevo quantity is then directly expressed through the symplectic eigenvalues as:
\begin{equation}\label{holevo1}
\chi_{BE}=G\bigg(\frac{\lambda_1-1}{2}\bigg)+G\bigg(\frac{\lambda_2-1}{2}\bigg)-G\bigg(\frac{\lambda_3-1}{2}\bigg),
\end{equation}
where $G(x)=(x+1)\log (x+1)-x\log x$ is the bosonic entropic function \cite{measures}. Thus, a lower bound on the key rate can be estimated. Unfortunately, the expression for the case of collective attacks is too lengthy to be presented in the paper and the corresponding calculations were performed numerically.

Similarly to the entanglement and purity, the positivity of the key rate (being the prerequisite for secure 
key distribution) appears to be sensitive to interdependence of the mean values of $\{\eta_i\}$, in 
particular, to the variance $Var(\sqrt{\eta})$, which limits the security of the scheme. 

Security breaking due to fading can already be observed in the case of individual attacks \cite{cvqkd1}, 
were the maximum value of $Var(\sqrt{\eta})$  can be obtained analytically:
\begin{equation}
Var(\sqrt{\eta})_{max,ind}=\frac{\langle\sqrt{\eta}\rangle^2\sigma-2(\sigma+1)(\chi+1)+\sqrt{\langle\sqrt{\eta}\rangle^4\sigma^2+4(\sigma+1)^2}}{2\sigma(\sigma+1)}.{} 
\end{equation}
Here, as mentioned, $\sigma=V-1$ corresponds to the variance of coherent state modulation in the prepare-and-measure scenario and will be used in the 
security considerations as the measure of modulation depth. 

Similar behavior is observed in the case of collective attacks. Moreover, since Gaussian excess noise 
$\epsilon_{f}$ caused by fading is state variance dependent, it appears that for 
each $\langle \sqrt{\eta} \rangle$ there is an optimal value of modulation $\sigma$, which maximizes 
the tolerable fading variance $Var(\sqrt{\eta})$. The corresponding plot of maximal variance $Var(\sqrt{\eta})$ 
versus modulation $\sigma$ is given in figure \ref{maxvar} (right) for different values of $\langle \sqrt{\eta} \rangle^2$. 
One can see, that the stronger the attenuation is, the lower must be the applied modulation, in order to minimize 
the negative impact of fading. If an additional untrusted noise $\chi$ is present, 
then the maximal $Var(\sqrt{\eta})$ is drastically reduced in the regime of low modulation $\sigma$, 
even under the amount of untrusted noise, which is negligible for high modulation.

Thus, in order to restore the security of CV QKD upon fading channels either upon high modulation variance or in the presence of
even a small amount of untrusted noise, the variance $Var(\sqrt{\eta})$ must be minimized. 
In practice, this could be achieved by demanding experimental efforts such as distillation \cite{distillation} which is aimed 
at restoring the state prior to the measurement, or by the use of PS, when a subset of the 
channel transmittance distribution $\{\eta_i\}$ are selected to contribute to the statistical ensemble of the post-selected state and to the key data, in case of the QKD. 
From the security point of view this can be done in the already stated assumption, that Alice and Bob are able to estimate 
the actual channel transmittances $\{\eta_i\}$, i.e. the estimation and information transmission is faster than the channel fluctuations. 
Thus, they post-select the sub-channels, which construct a new state with the covariance matrix of the form (\ref{evolvedstate}), and all 
Gaussian security analysis based on the extremality of Gaussian states and optimality of Gaussian attacks are valid for 
the new set of sub-channels and the new resulting state. 

In the further sections we consider a particular theoretical model of a fading channel and then apply our results to the 
experimentally obtained distributions.

\section{Fading due to beam wandering}
\label{beam-wandering}

\begin{figure}[h]
\centerline{\begin{tabular}{ll}
\includegraphics[width=0.35\textwidth]{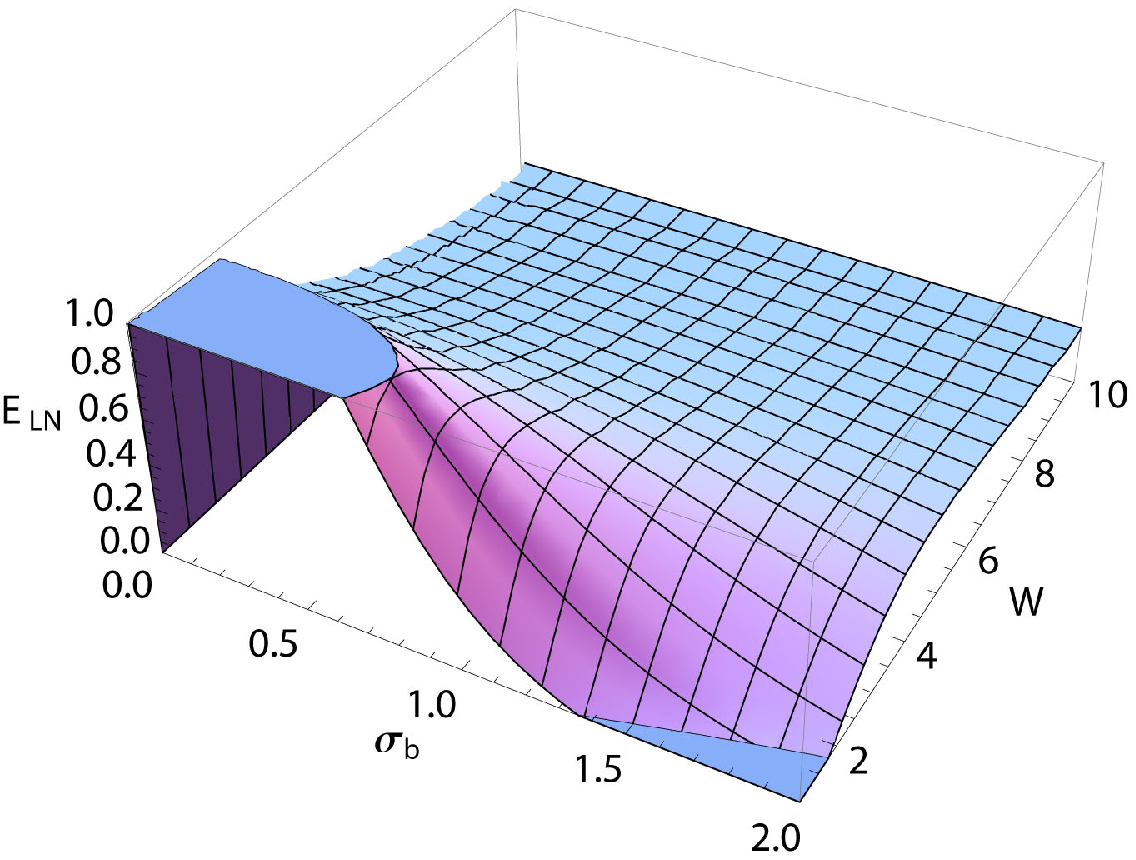}
\includegraphics[width=0.35\textwidth]{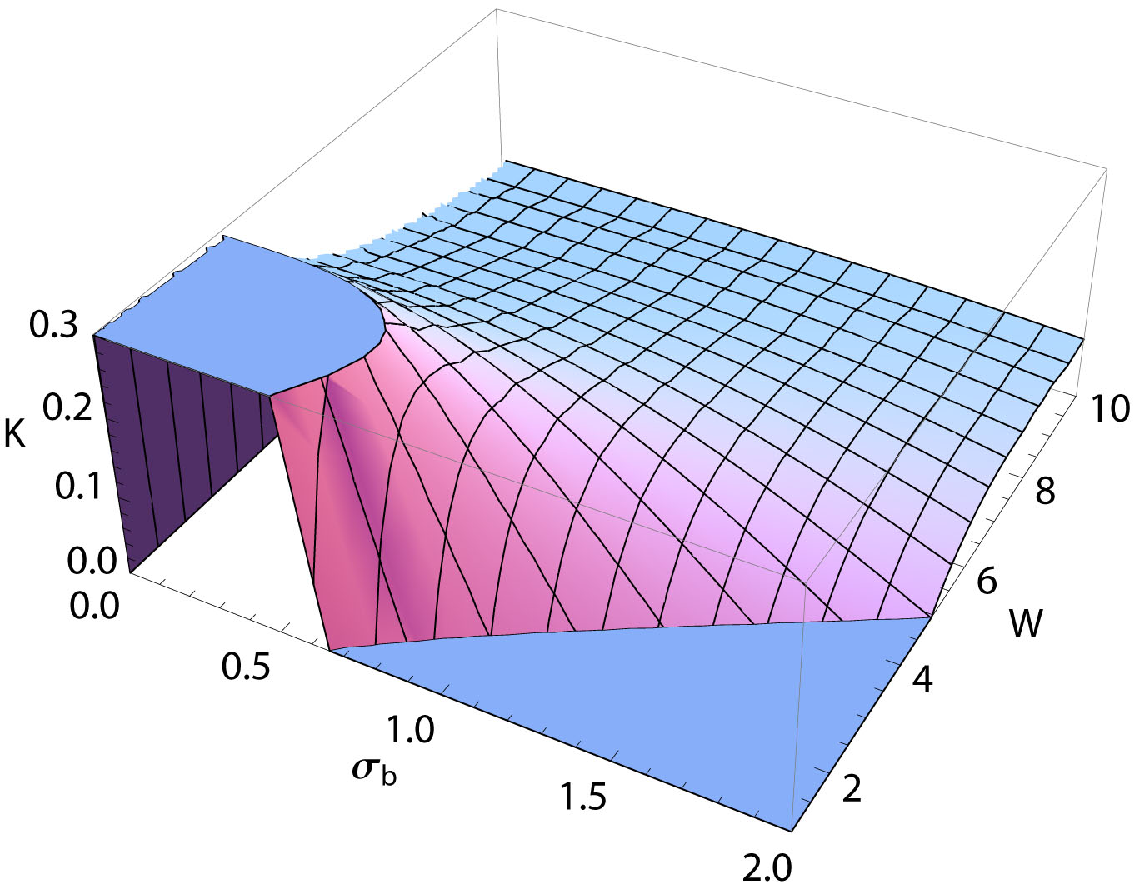}	
\end{tabular}}
\caption{Left: the logarithmic negativity $E_{LN}$ of the initially pure entangled state after a fading channel; right: key rate secure against collective attacks (in bits per measurement) for the initially pure entangled state  after a fading channel. The state variance $V=10$, the channel is characterized by the beam position variance $\sigma_b$ and the spot size $W$ related to the aperture size $a=1$. 
\label{3dmodels}}
\end{figure}

In realistic atmospheric channels, the main impact of turbulence is beam wandering, when the beam spatially fluctuates around the receiver's aperture.
Fading due to beam wandering is well described by a recently obtained statistical 
distribution \cite{rostock}, which, in the case of beam alignment around the aperture center, is given by the log-negative 
Weibull distribution. The distribution is governed by the shape and scale parameters, expressed by beam-center position variance 
$\sigma_b^2$, beam-spot radius $W$, aperture radius $a$, and cut at the maximum transmission value, which is given by
\begin{equation}
\label{eta0}
\eta_0^2=1-\exp\big({-2\frac{a^2}{W^2}\big)}.
\end{equation}
Here we numerically analyze the entanglement properties of the Gaussian states and the security of the respective CV QKD protocols over the fading channels, governed by beam wandering. 
First, we show how the logarithmic negativity behaves with respect to beam 
position variance $\sigma_b$ and spot size $W$ (normalized by the dimensionless aperture size $a=1$, as the fading due to beam wandering depends on the ratio $a/W$), the results are given in figure \ref{3dmodels} (left). Then, we estimate the key rate, secure against collective attacks, in 
the same fading channel assuming pure channel loss and no untrusted detection noise (results are given in figure \ref{3dmodels}, right). It is evident, 
that both entanglement and security are sensitive to the variance of beam position fluctuations. 
At the same time, an increase of the beam spot radius is able to partly compensate its negative effect (which is already
known from classical communications \cite{wandering}), 
restoring security and Gaussian entanglement, but limiting the values of the key bits and logarithmic 
negativity obtained from the covariance matrix.

Now let us assume that PS of sub-channels is applied by reducing the region of transmittance values for which the data, contributing to the resulting state and to the key, are taken. We define each sub-channel $\eta_i$ as the set of events, in which the transmittance was relatively stable and lies within the region of width $\Delta \eta$ centered around $\eta_i$. By the stability of transmittance, we mean that the transmittance fluctuations within the region $\Delta \eta$ are negligible (fractions of maximum variance, shown in figure \ref{maxvar}). Thus, the set of sub-channels corresponds to the uniform binning of the continuous transmittance distribution with a bin size of $\Delta \eta$. In the simplest case of normally distributed transmittance values, the binning effect on the second moments would be described by Sheppard`s correction \cite{binning}, while for the Weibull distribution, cut at $\eta_0$ (\ref{eta0}), the correction caused by a uniform binning changes unevenly due to the cut (i.e. binning must be optimized). In any case, if the number of bins is larger than 1, the second moment of the discretized distribution is always larger than that of an actual continuous distribution, thus being a pessimistic assumption and complying well with the security considerations. The detailed study of the effect of fading discretization on security at practically achievable estimation precision will be the subject of future research, while in the following we take $10^3$ bins for the theoretical analysis and $10^2$ bins for the experimental data as a good although pessimistic approximation of the channel properties.

\begin{figure}[h]
\centerline{\includegraphics[width=0.7\textwidth]{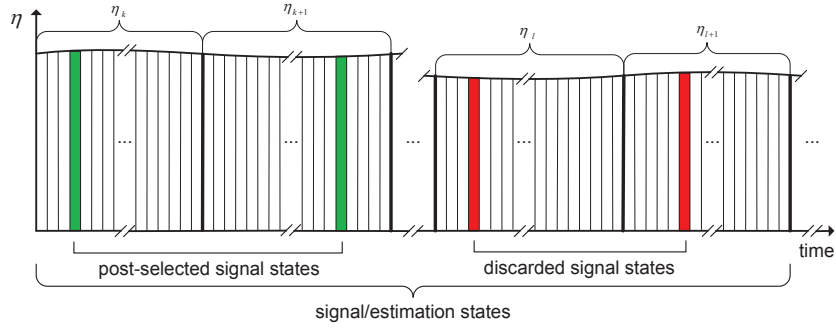}}
\caption{Time-flow of the Gaussian CV QKD protocol in a fading channel with PS. The slowly variable transmittance $\eta$ is divided into sub-channels of width $\Delta \eta$ centered around $\{\eta_i\}$, when the transmittance lies within the region $\eta \in [\eta_i \pm \Delta \eta/2]$ and is considered approximately stable. Multiple quantum states are sent through the channel during its stability time, thus corresponding to a given sub-channel, with the number of pulses being large enough for the proper estimation of sub-channel properties. Then, a few states (at least one per each occurrence of a given sub-channel) are randomly chosen to carry the signal (the events depicted as filled), thus being indistinguishable from the estimation states. Further, depending on the PS region, the results of the signal states measurements either are discarded (sub-channels $\eta_l,\eta_{l+1}$, events in red) or contribute to the resulting post-selected covariance matrix, used for security analysis, and to the raw key data (sub-channels $\eta_k,\eta_{k+1}$, events in green).  
\label{fading_post_selection_timeflow}}
\end{figure}

The security assumptions imply that the trusted parties know the effect of the channel on the signal states and are able to reconstruct the corresponding two-mode covariance matrix. The proper sub-channel estimation requires a large number of estimation states to be sent through the channel during its stability. The estimation states must be indistinguishable from the signal states so that an eavesdropper could not affect the estimation and signal states separately. Then, some of the states (at least one) for each sub-channel occurrence are randomly chosen as signal ones to contribute to the raw key and the corresponding covariance matrix. Thus, the security is not shown for each sub-channel occurrence separately (as it would require an asymptotic number of signal states combined with multiple estimation pulses sent upon each such event). Instead, the security is shown for the state, being conditionally prepared from the set of data obtained upon the post-selected sub-channels. The schematic time-flow of the Gaussian CV QKD protocol in a fading channel with PS is given in figure \ref{fading_post_selection_timeflow}, illustrated by the set of estimation and randomly chosen signal pulses upon each sub-channel occurrence.

Let us first consider PS of a single sub-channel and estimate entanglement per state and the amount of secret key per state (not taking into account the overall success probability of PS). The maximum key and maximum entanglement per state will then be achieved for a fixed non-fluctuating channel with the highest transmission possible,
given by (\ref{eta0}). Thus the ratio between the beam spot and the aperture size must be maximized if the maximum secure 
information per measurement is required, regardless of success probability and, accordingly, transmission time. 

\begin{figure}[h]
\centerline{\includegraphics[width=0.5\textwidth]{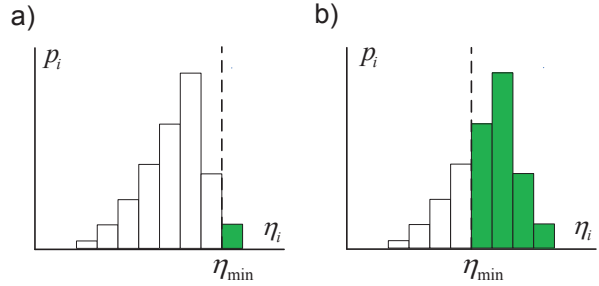}}
\caption{Scheme of PS of sub-channels. The data is accumulated upon different transmittance 
values, which lay within the sub-channel transmittance regions. Then only data from the post-selected 
maximally transmitting non-zero probability sub-channel (a, filled) or several sub-channels (b, filled) are
selected to contribute to the post-selected state and the respective key whose security is derived from the covariance matrix of the post-selected state.
\label{fading_post_selection}}
\end{figure}

The principle of PS of a maximally transmitting sub-channel is schematically depicted in figure \ref{fading_post_selection} (a). 
The data are accumulated in different sub-channel transmittance regions, in which the transmittance fluctuations are assumed to be negligible. Then only the data from the post-selected maximally transmitting sub-channel is chosen to contribute to the post-selected state and the respective two-mode covariance matrix. 

However, if the overall success probability is taken into account, then the PS region must be extended (schematically depicted in figure  \ref{fading_post_selection}, (b)) in order to provide the optimal trade-off between the probability for transmittance to fall within the post-selected region and the fading variance of the post-selected sub-channels, reducing the secure key rate. 

Let us consider the effect of the single most transmitting sub-channel PS on the entanglement properties of Gaussian states and security of the respective CV QKD protocol. 

We access the available overall entanglement (beyond the Gaussian one) by estimating the conditional entropy of the state after PS, 
which corresponds to the lower bound on the distillable entanglement \cite{eisert}:
\begin{equation}
E_D > S(\rho_{A})-S(\rho),
\end{equation}
where $S(\cdot)$ denotes the von Neumann entropy of the state, the density matrix $\rho$ refers to the state after 
the fading channel and PS and $\rho_A$ refers to the reduced state with respect to sub-system A. 
The entropies are directly calculated from the symplectic eigenvalues of the respective covariance matrices. 
Thus, we can check whether entanglement of the state is restored by confirming a non-zero conditional entropy as well as
quantify the amount of entanglement by the PS. The logarithmic negativity cannot be used for this purpose 
since it represents an upper bound on the entanglement.  

\begin{figure}[h]
\centerline{\begin{tabular}{ll}
\includegraphics[width=0.35\textwidth]{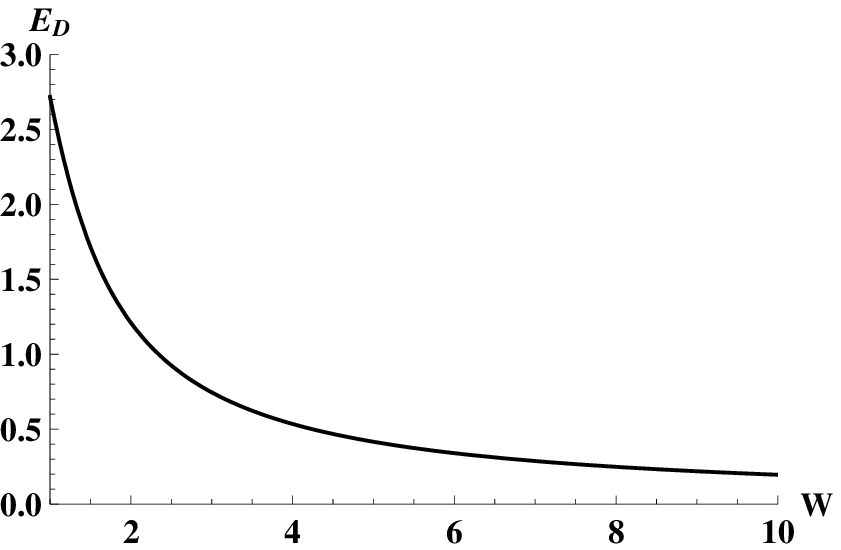}
\includegraphics[width=0.35\textwidth]{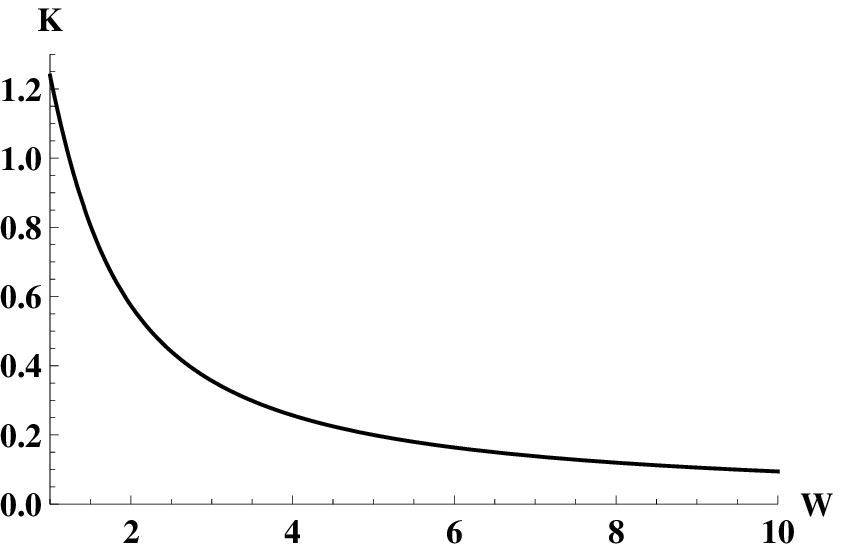}	
\end{tabular}}
\caption{The maximum conditional entropy (left) and the maximum lower bound on the key rate (right, in bits per measurement) for the post-selected 
maximally transmitting sub-channel with respect to the spot size $W$ (normalized by $a=1$) upon initially pure entangled 
state with variance $V=10$.
\label{3dmodelsPS}}
\end{figure}

The results of the calculations are given in figure \ref{3dmodelsPS} for the conditional entropy (left) and the key rate (right). 
It is evident that PS of the highest transmitting sub-channel increases the entanglement and security 
properties of the state as well as removes the dependence on the beam fluctuations variance, as the maximum 
available transmission (\ref{eta0}) does not depend on the variance.

\begin{figure}[h]
\centerline{\begin{tabular}{ll}
\includegraphics[width=0.35\textwidth]{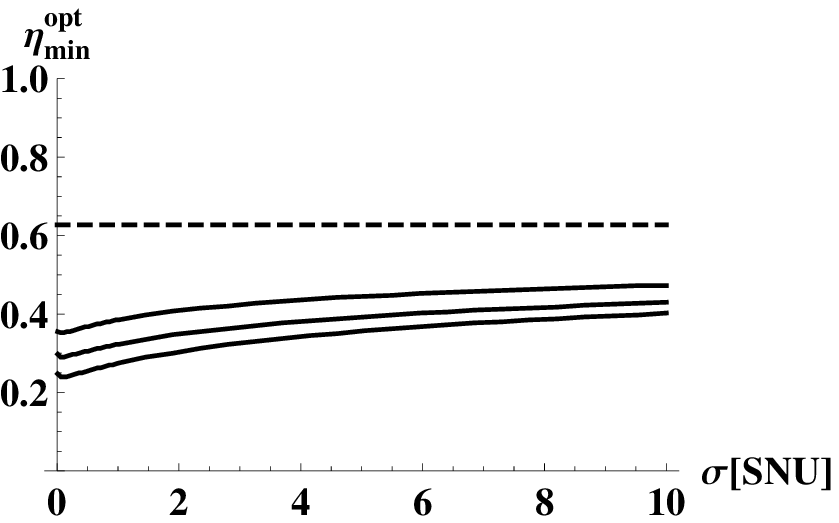}
\includegraphics[width=0.35\textwidth]{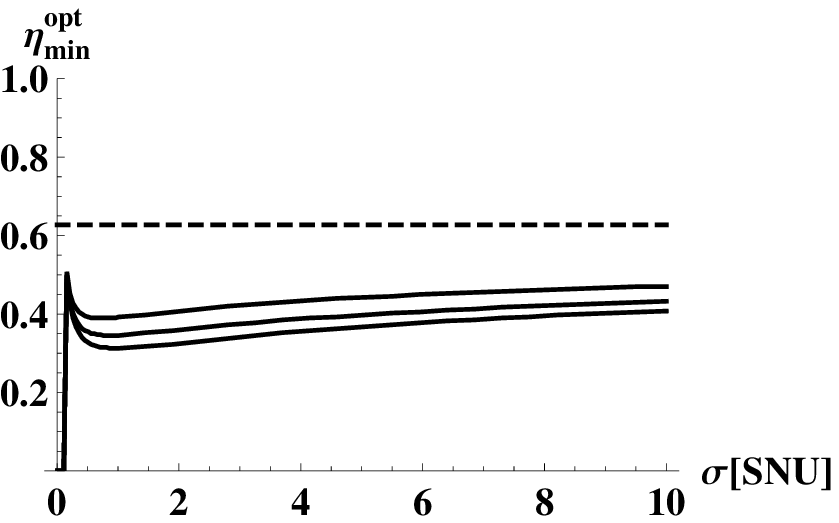}	
\end{tabular}}
\caption{The beginning of the optimal PS region versus the effective state modulation $\sigma=V-1$ upon no untrusted noise (left) and in the presence of untrusted noise $\chi=0.01$ (right). The horizontal dashed line represents the cut of the transmittance distribution, given by (\ref{eta0}). Beam wandering variance $\sigma_b$ is $0.5,1,3$ (from top to bottom), aperture size $a=1$, beam spot size $W=2$.
\label{start_opt}}
\end{figure}

However the key rate essentially depends on the overall success probability of the PS. Thus, we weight the post-selected key rate by the overall probability for transmittance to fall within the PS range. Such a weighted key now can be maximized by the optimal choice of the PS region, i.e. of the boundary $\eta_{min}$ between the selected and neglected sub-channels (see figure \ref{fading_post_selection} (b)), which (for a typical single-peaked fading distribution) explicitly describes the optimal PS region. The position of such a boundary depends on the effective modulation $\sigma=V-1$ and untrusted excess noise, as well as on the parameters of the set-up. We plot the typical dependence of the optimal PS region boundary with respect to the state modulation in figure \ref{start_opt} for the case of no untrusted excess noise present (left) and of additional untrusted noise $\chi=0.01$ (right) upon different values of beam wandering variance $\sigma_b$ for the beam spot being twice bigger than the aperture size: $W=2$. The beginning of the PS region $\eta_{min}$ is compared with the cut of the transmittance distribution, given by (\ref{eta0}). Thus, the offset between solid and dashed lines illustrates the width of the optimal PS region for different modulation depths and beam wandering variances.

Here we are interested in restoring security when it was lost due to fading, while quantitative estimations will be done in the next sections. It is evident from the graph, that if no untrusted noise is present, the optimal PS region shrinks continuously with an increase of modulation, i.e. stronger PS must be optimally applied for highly modulated states. However, even if only a small fraction of untrusted noise is present, the protocol, based on the Gaussian states over a fading channel, first becomes unsecure upon any PS (when $\sigma$ is low), then strong PS must be applied, as weakly modulated states are more sensitive to the combination of fading and untrusted noise and then as the modulation increases, the PS can be first released and then made more strict upon further increase of effective modulation. 

In the following section, we apply our analysis to the particular experimentally obtained distribution, which is well fitted by the beam wandering and fixed attenuation.

\section{Real fading channel}
\label{real-channel}

We check the above given considerations for a real short-distance atmospheric fading channel, where fading is mostly the result of beam wandering. 
We consider a particular experimental distribution, obtained by a relative intensity measurement over a free-space link. 

\begin{figure}[h]
\centerline{\includegraphics[width=0.8\textwidth]{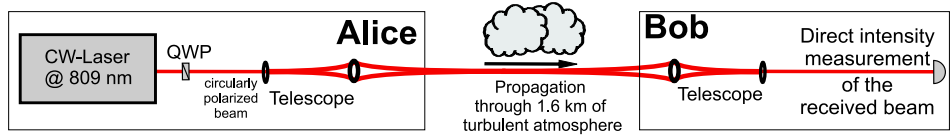}}
\caption{Experimental setup: A circularly polarized laser beam is sent through 
the 1.6 km free-space channel after having been expanded with the help 
of a telescope. At the receiver, the beam diameter is reduced again and 
its intensity is directly measured
\label{channel_setup}}
\end{figure}

The experimental data were recorded within the framework of a setup on 
free-space quantum communication using continuous polarization 
variables, see figure \ref{channel_setup} \cite{B1}. This experiment is a further 
development of earlier work performed in the laboratory \cite{B2} and on the 
roof of the MPL`s building in Erlangen \cite{B3,B4}. We use a grating-stabilized 
diode laser, with wavelength of 809 nm that lies within an atmospheric 
transmission window. The linearly polarized laser beam is transferred to 
a circularly polarized state and sent through the 1.6 km free-space 
channel from the institute's building to a tall university building, after having been expanded. Bob 
then reduces the beam diameter of approximately 120 mm by a telescope 
with an entrance aperture of 80 mm.

\begin{figure}[h]
\centerline{\includegraphics[width=0.35\textwidth]{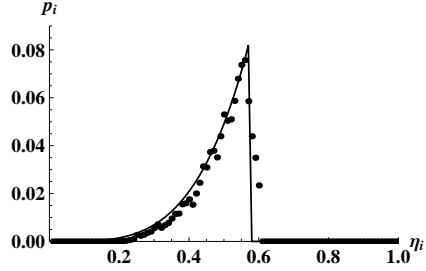}}
\caption{Distribution of transmittance ${\eta_i}$ through the atmospheric fading 
channel (dots) as well as fitted log-negative Weibull distribution \cite{rostock}  
(solid line). The intensity of the received beam was directly measured 
at a sampling rate of 150kHz. The figure shows a histogram of the channel transmission for 10s of experimental data with a bin size of $\Delta \eta = 0.01$. 
The intensity of the outgoing beam was 12 mW and the mean 
value of the received beam 5.92 mW, with a standard deviation 0.92 mW.
\label{distribution}}
\end{figure}

For the scheme mentioned in \cite{B1}, this beam will serve as a carrier beam for quantum information modulated onto its polarization degree of freedom.
Here, on the other hand, in order to characterize the transmission 
properties of the fading atmospheric channel, we perform a direct 
detection of the carrier beam power. The  data, normalized by the outgoing beam power at Alice, are shown in the histogram in figure \ref{distribution}, 
along with a fit according to the log-negative  Weibull distribution \cite{rostock}.
The experimental data match very well the 
theoretical distribution given in the previous section with $\sigma_b=0.6, W/a=1.5$ and
upon an additional fixed attenuation of $25\%$ (given by fixed losses at optical components). 

\begin{figure}[h]
\centerline{\begin{tabular}{ll}
\includegraphics[width=0.35\textwidth]{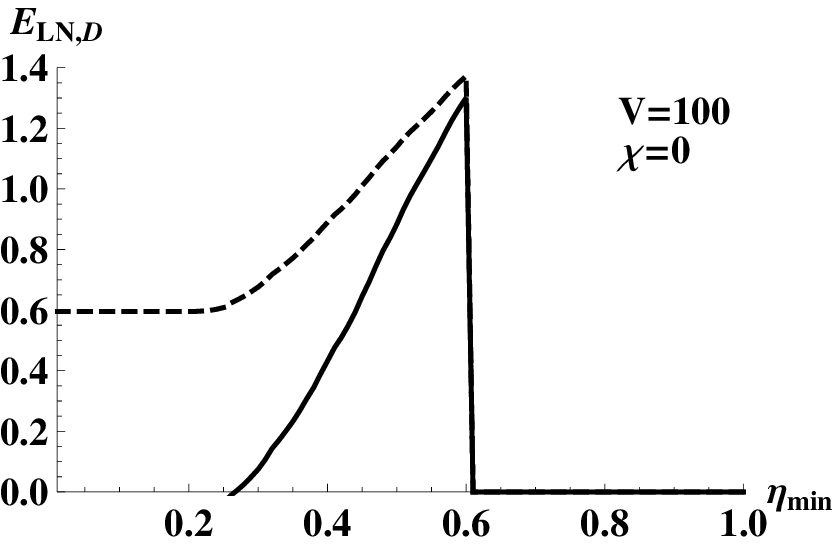}
\includegraphics[width=0.35\textwidth]{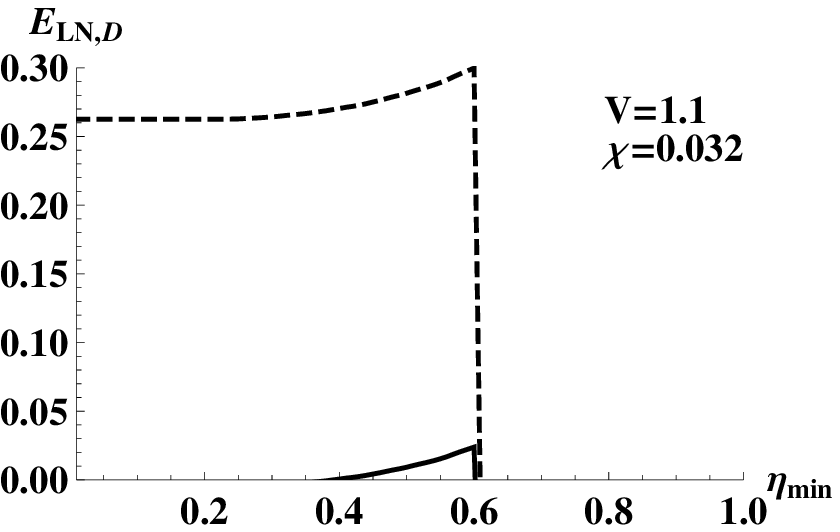}
\end{tabular}}
\caption{Effect of PS (represented by the beginning of PS region) after the fading channel 
given in figure \ref{distribution} on the entanglement of the states in terms of logarithmic negativity (dashed line)
and conditional entropy (solid line) for high state variance $V=100$ (left) and low variance $V=1.1$ (right) in the
presence of noise $\chi=3.2\cdot 10^{-2}$.
\label{entPS}}
\end{figure}

Let us first analyze the effect of PS on the entanglement properties of the states after passing through the experimental channel. Estimation of 
entanglement with respect to PS in terms of the beginning of the PS region (ending at $\eta_i=1$) was performed numerically and is given in figure \ref{entPS}.

The channel shown in figure \ref{distribution} appears to be entanglement preserving in terms of logarithmic negativity. 
However, the lower bound on distillable entanglement (given by the conditional entropy) is below 0 without PS 
(which is due to the fact that distillable entanglement is more sensitive to noise induced by fading for high state variance or to additional noise for low variance), but can be restored by PS, which thus guarantees the increase of entanglement in the resulting state, especially upon higher state variance.

\begin{figure}[h]
\centerline{\begin{tabular}{ll}
\includegraphics[width=0.35\textwidth]{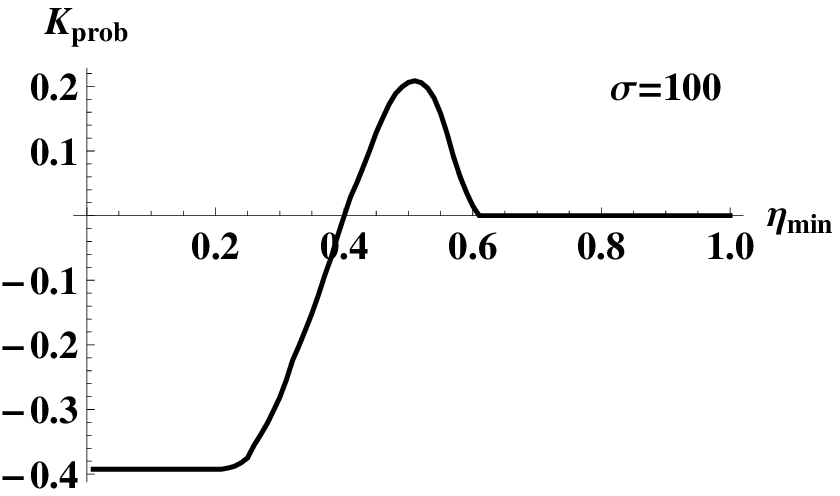}
\includegraphics[width=0.35\textwidth]{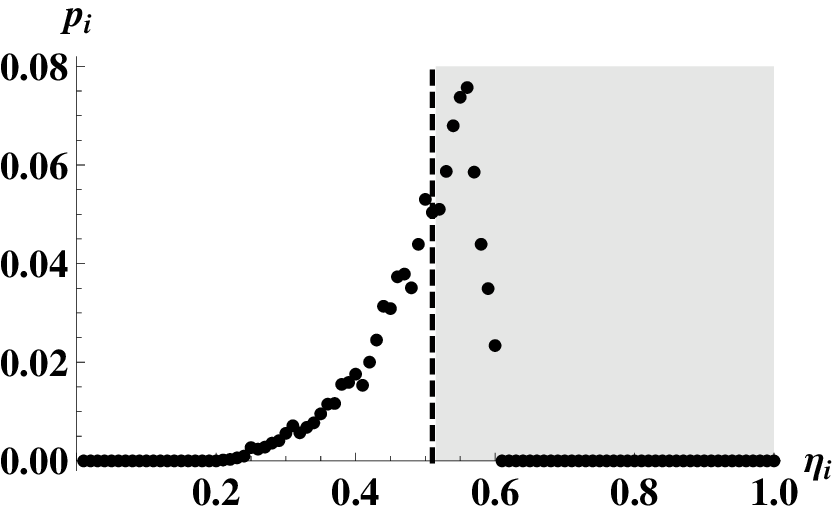}
\end{tabular}}
\centerline{\begin{tabular}{ll}
\includegraphics[width=0.35\textwidth]{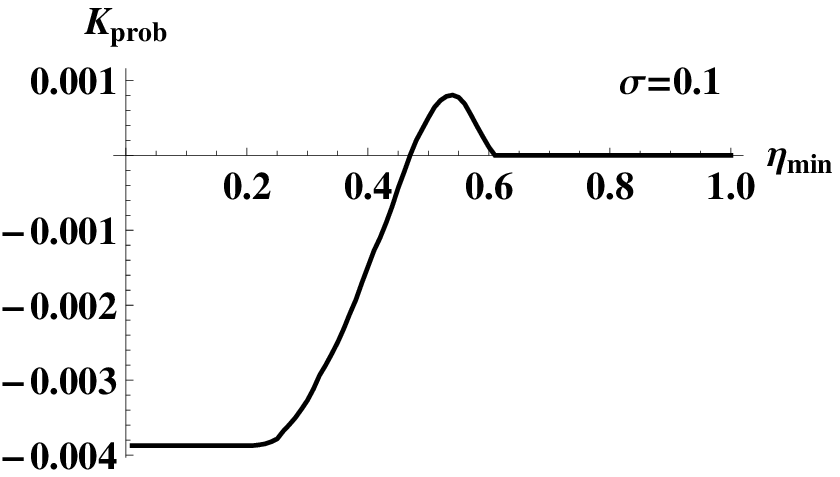}
\includegraphics[width=0.35\textwidth]{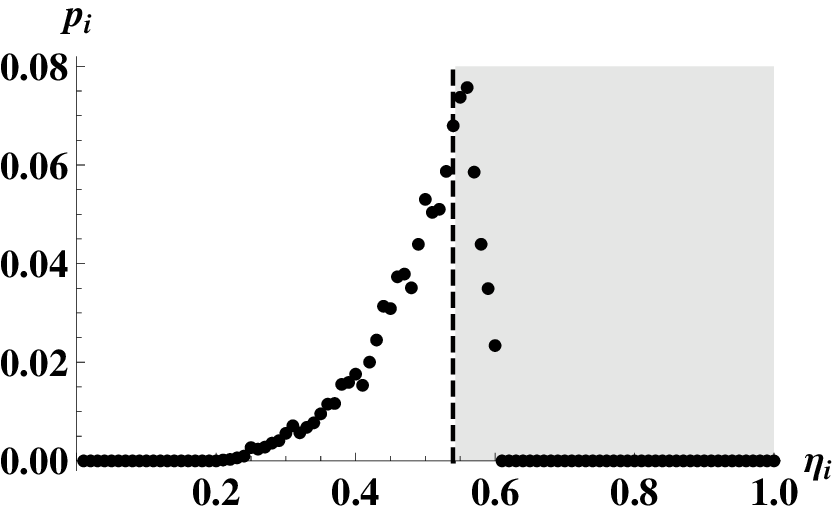}
\end{tabular}}
\caption{Left: effect of PS (represented by the start of the PS region) after the fading channel 
given in figure \ref{distribution} on the security of the states against collective attacks in terms of the lower bound 
on the key rate (in bits per measurement), weighted by PS success probability, for high modulation variance $\sigma=100$ (up) and low modulation variance $\sigma=0.1$ 
in the presence of noise $\chi=0.032$. Right: respective optimal PS region (bounded by the corresponding $\eta_{min}$), maximizing the weighted key rate for given parameters, indicated 
atop of the experimentally obtained fading distribution.
\label{keyPS}}
\end{figure}

Security against collective attacks is also sensitive to fading in the channel given in figure \ref{distribution}, but restorable with PS. 
The channel, being characterized by $\langle \sqrt{\eta} \rangle^2 \approx 0.492$ and 
$Var(\sqrt{\eta}) \approx 3\cdot10{^-3}$, appears to be Gaussian CV QKD security breaking either upon high modulation variance 
$\sigma=100$ (which, for given $\langle \sqrt{\eta} \rangle^2$, can tolerate $Var(\sqrt{\eta})$ not 
larger than $1.2\cdot 10^{-3}$) or in the presence of untrusted noise. However, PS is able to 
restore security and must be optimized with respect to the key rate, weighted by PS success probability. This is evident from 
the graphs of the weighted key rate versus the PS region given in figure \ref{keyPS} for both high and low modulation variances 
along with the optimal PS regions for the given modulation. 
Here we take into account the overall success probability (i.e. the probability for channel transmission to appear 
in the post-selected region) and weight the key rate by this success rate, which requires optimization of the PS 
region to maximize the weighted key.

Here we assume that the estimation is reliable and is not affected by the PS. Investigation of Eve's possible impact on estimation efficiency 
(which can seriously affect the security analysis already for non-fading channels) is beyond the scope of this paper and will be the subject 
of future work. 

\begin{figure}
\centerline{\psfig{width=8.0cm,angle=0,file=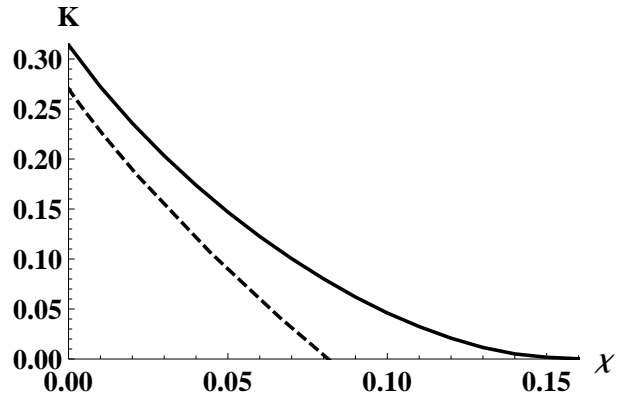}}
\caption{Secure key rate $K$ (in bits per measurement) over the fading channel described by the experimentally obtained distribution (given in figure \ref{distribution}) for the given untrusted excess noise $\chi$ (in SNU) upon optimized PS 
and optimal modulation $\sigma$ (solid line) and without PS, upon optimal modulation $\sigma$ (dashed line).}
\label{optimization}
\end{figure}

Evidently, there exists a trade-off between robustness of the protocol to fading (provided by reduced modulation)
and robustness to noise (provided by increased modulation and requiring optimal PS). To generalize the 
result, we performed an optimization of both the state modulation $\sigma$ and the PS region, for a given untrusted 
excess noise $\chi$, modeled in the experimentally measured channel, and compared the resulting key rate to that with optimized modulation $\sigma$, but no PS. 
The corresponding plot is given in figure \ref{optimization}. The effect of PS is clearly visible from 
the graphs. In the presence of untrusted excess noise above the threshold (being approximately $0.08$ SNU for 
the given distribution), no modulation can provide security over the given channel without PS. 
Contrary to this, if the optimized PS is performed and the modulation is also optimal, the scheme is 
able to tolerate almost double the amount of untrusted noise.

\section{Effects of a limited ensemble size for the key rate and channel estimation}
\label{finite-size}

The security considerations above were performed in the standard assumption of the asymptotic limit, when
the size of the data ensemble, contributing to the key rate, is assumed infinite. However, this is never the case in 
practice and consideration of finite-size effects is essential for the CV QKD analysis with the first steps in this direction
already being performed in \cite{finite-size}. The issue becomes even more important when PS is applied as it 
reduces the ensemble size. Thus, in this section we briefly analyze the statistical properties of the lower bound on the 
key rate with respect to the ensemble reduction, caused by PS, by checking the convergence of the key rate 
to the theoretical prediction depending on the PS range. 

\begin{figure}[h]
\centerline{\includegraphics[width=0.9\textwidth]{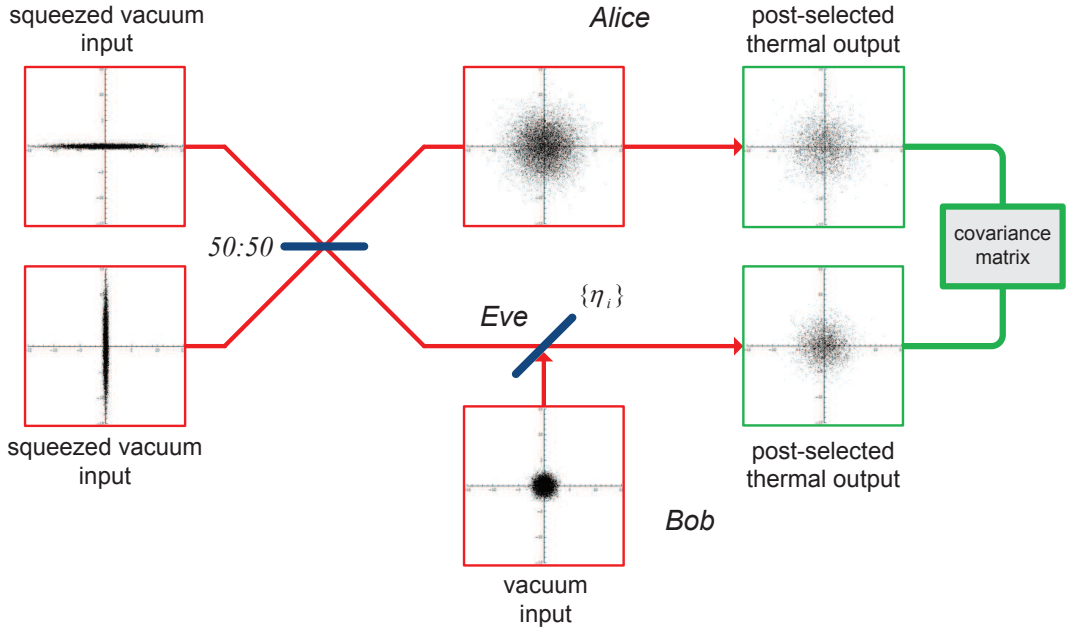}}
\caption{Scheme of numerical modeling for the analysis of the effects of the finite ensemble size on the reliability of PS. Two Gaussian number generators produce 
 distributions of two orthogonally squeezed vacuum states, which are then mixed with the coupling ratio of 50:50. One of the modes is coupled to the Gaussian number generator, emulating vacuum input of the purely lossy fading channel, the coupling ratio being generated according to the experimentally obtained transmittance probability distribution in a fading channel. The resulting state is post-selected (as on figure \ref{fading_post_selection}, (b)) in both modes according to the PS region. The two-mode covariance matrix is calculated and used for the security estimation. The depicted distributions were built for the source variance of $V=10$ (corresponding to squeezing parameter $r \approx 1.49$) and the PS region starting with $\eta_{min}=0.55$.
\label{fading_finite_scheme}}
\end{figure}

We perform a statistical simulation, introducing two Gaussian number generators, producing quadrature distributions of the
two orthogonally squeezed vacuum states. We mix these distributions using a beamsplitter with the input-output relation (\ref{inout}). We emulate coupling of the states on the $50:50$
beam splitter by setting $\eta=1/2$ to produce entangled states. The third Gaussian number generator produces the quadrature distributions of the
vacuum mode, which emulates a purely lossy channel and is similarly coupled to one of the modes. The value of transmittance is randomly 
picked for each next data point according to the distribution given in figure \ref{distribution}, i.e. the worst case assumption of the ability to transmit only a single bit 
through the given estimated channel is made. After coupling, we obtain the ensemble of the quadrature values
of the three modes after the channel transmission. Then we post-select the data according to a certain region and calculate the variance and covariances between the modes to construct 
the covariance matrix of elements $\gamma_{ij}=\langle r_ir_j\rangle-\langle r_i \rangle\langle r_j \rangle$ ($r=\{x,p\})$. 
The scheme of the numerical modeling procedure is given in figure \ref{fading_finite_scheme}.

\begin{figure}[h]
\centerline{\begin{tabular}{ll}
\includegraphics[width=0.35\textwidth]{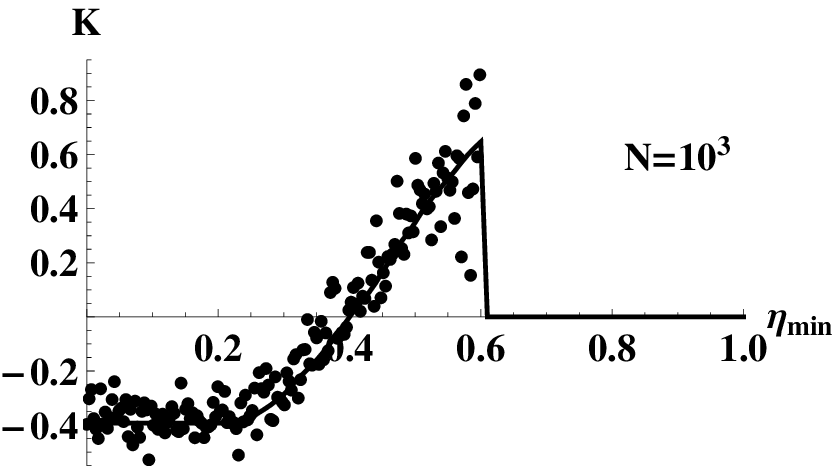}
\includegraphics[width=0.35\textwidth]{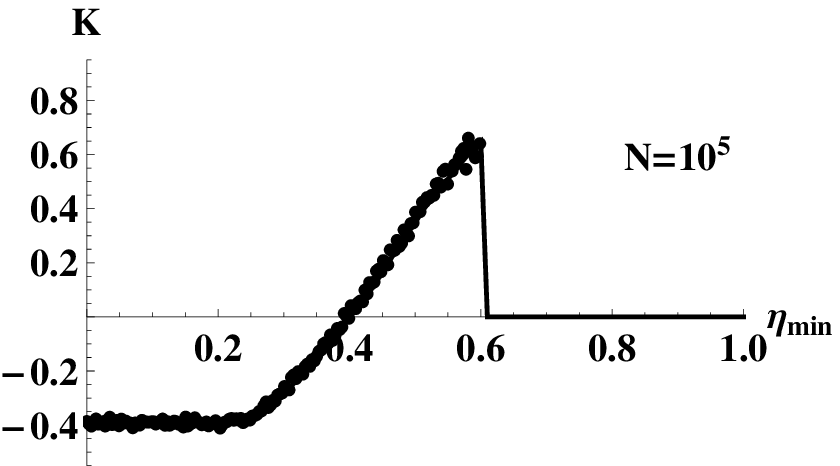}
\end{tabular}}
\caption{Effect of the finite ensemble size on the post-selected key rate upon high modulation variance $V=100$ with respect 
to the beginning of the PS region for initial ensemble size of $10^3$ (left) and $10^5$ (right) data points.
\label{finitesize}}
\end{figure}

The obtained covariance matrix is checked to comply with the uncertainty principle: $\gamma+i\Omega \geqslant	0$ (where $\Omega$ is the symplectic form) \cite{measures}
and the physically valid covariance matrix is then used to estimate the key rate, using the above described Gaussian machinery, in assumption of purification performed by Eve (as Eve's mode is traced out in the final covariance matrix). The resulting key rate for the given ensemble size can then be compared to the key rate expected in the asymptotic limit. The results of the modeling are given in figure \ref{finitesize} for different initial ensemble sizes along with the 
theoretical prediction in the asymptotic regime. 

It is evident from the graph, that while an ensemble size of $10^3$ gives strongly fluctuating results and is sensitive to PS,
which additionally increases the fluctuations, an ensemble size of $10^5$ already gives good convergence with the theoretical
prediction of the asymptotic limit with some divergence upon the most tight PS. Thus, our previous analysis, performed in 
the asymptotic regime is valid upon finite ensemble size (assuming perfect estimation) and the positive effect of
PS is confirmed in the finite-size regime (provided the estimation is still precise).

\begin{figure}[h]
\centerline{\includegraphics[width=0.35\textwidth]{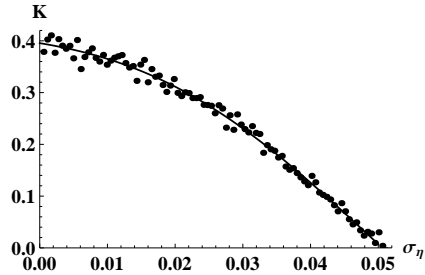}}
\caption{Effect of the imperfect estimation, modeled as the normal distribution of actual transmittance values with variance $\sigma_{\eta}$ around the "estimated" values, corresponding to the distribution in figure \ref{distribution}, on the optimally post-selected key rate at a high modulation $V=100$ and an ensemble size of $10^5$ points. The results of numerical modeling (dots) are given along with the fitting curve (solid line).
\label{KR_vs_eta_fluctuations}}
\end{figure}

We also address the effect of imperfect estimation (which can also be the result of the finite number of signal pulses) on the efficiency of PS. In the context of our finite-size modeling we now assume that the actual values of channel transmittance are normally distributed with variance $\sigma_{\eta}$ around the transmittance, picked according to the experimentally obtained distribution (see figure \ref{distribution}). The effect of such a randomization on the optimally post-selected key rate for a high modulation variance $V=100$ is given in figure \ref{KR_vs_eta_fluctuations}. The results show the stability of PS improvement against the imperfect estimation, which breaks the security of the protocol for a transmittance measurement variance of about $0.05$. The number of data points for the numerical analysis is taken $10^5$ and still causes some visible fluctuations. States with the weaker modulation appear to be more robust against imperfect estimation, which thus has the similar effect as fading. Moreover, such a transmittance randomization due to imperfect estimation appeared to be more threatening to security than worst-case assumptions for transmittance to be constantly decreased by $3\sigma_\eta$ or $6\sigma_\eta$. This is due to the fact that fading in certain conditions more strongly affects security than the reduction of constant attenuation.

\section{Summary and conclusions}
We investigated the effect of fading channels on the entanglement properties of Gaussian states and 
the security of coherent states-based CV QKD and showed the sensitivity of entanglement, purity and
security of the scheme to 
the interdependence of transmittance mean values $\langle \sqrt{\eta} \rangle$ and $\sqrt{\langle \eta \rangle}$. 
The effect of a fading channel is equivalent to state variance-dependent untrusted excess noise. 
Thus, weakly modulated states are less sensitive to the statistical properties of the transmittance distribution. 
However, the presence of further untrusted excess noise reduces the tolerance of weakly entangled states to fading.
Moreover, for a short-distance fading channel, where fluctuations of transmittance are caused by beam wandering, the variance
of the spot distribution is more destructive to entanglement and security, while an increase of the spot size with respect to 
aperture can partly compensate for its negative effect. 

PS of sub-channels (in the assumption of their proper estimation) 
was shown to be able to restore distillable entanglement and security for both high and low modulation upon noise, and should be the 
subject for optimization, as shown for the data from the real atmospheric channel. 
The combination of optimized modulation and optimal PS allows transmission with almost twice as high 
untrusted noise. The positive effect of PS on QKD security is also confirmed for a finite ensemble size when statistical convergence 
of the key rate is tested and requirement of the effective transmittance estimation is addressed. The proposed sub-channel PS method thus can 
be used as a feasible alternative to the experimentally demanding entanglement distillation protocols if fading channel parameters are properly estimated.

\ack
VCU was supported by project no. P205/10/P321 of the Grant Agency of Czech Republic and acknowledges valuable discussions with Norbert L\"utkenhaus. BH gratefully acknowledges the funding of the Erlangen Graduate School in Advanced Optical Technologies (SAOT) by the German Research Foundation
(DFG) in the framework of the German excellence initiative. RF acknowledges support from the Alexander von Humboldt foundation. In addition, the project was supported under FP7 FET Proactive by the integrated project Q-Essence and CHIST-ERA (Hipercom).

\end{document}